\documentclass[12pt,preprint]{aastex}

\usepackage[english]{babel}
\usepackage[latin1]{inputenc}
\usepackage{xspace}

\newcommand{\sn}{SN\xspace}
\newcommand{\sne}{SNe\xspace}
\newcommand{\snia}{SN~Ia\xspace}
\newcommand{\sneia}{\sne~Ia\xspace}

\def\lsim{\raise0.3ex\hbox{$<$}\kern-0.75em{\lower0.65ex\hbox{$\sim$}}}
\def\gsim{\raise0.3ex\hbox{$>$}\kern-0.75em{\lower0.65ex\hbox{$\sim$}}}

\shorttitle{Lensed SN at $z=1.703$ behind A1689}
\shortauthors{Amanullah et al.}

\begin{document}


\title{A highly magnified supernova at $z=1.703$ behind the massive
  galaxy cluster A1689}


\author{%
  R.~Amanullah\altaffilmark{1,2},
  A.~Goobar\altaffilmark{1,2},
  B.~Cl\'ement\altaffilmark{3},
  J.-G.~Cuby\altaffilmark{3},
  H.~Dahle\altaffilmark{4,5},
  T.~Dahl\'en\altaffilmark{6},
  J.~Hjorth\altaffilmark{7},
  S.~Fabbro\altaffilmark{8},
  J.~J\"onsson\altaffilmark{1,2},
  J.-P.~Kneib\altaffilmark{3},
  C.~Lidman\altaffilmark{9},
  M.~Limousin\altaffilmark{3},
  B.~Milvang-Jensen\altaffilmark{7}
  E.~M\"ortsell\altaffilmark{1,2},
  J.~Nordin\altaffilmark{1,2},
  K.~Paech\altaffilmark{10},
  J.~Richard\altaffilmark{11,7},
  T.~Riehm\altaffilmark{12,2},
  V.~Stanishev\altaffilmark{13},
  D.~Watson\altaffilmark{7}
}
\altaffiltext{1}{Department of Physics, Stockholm University, 
  Albanova University Centre, SE~106-91 Stockholm, Sweden; rahman@fysik.su.se} %
\altaffiltext{2}{The Oskar Klein Centre, Physics Department, 
  Stockholm University,
  Albanova University Centre, SE~106-91 Stockholm, Sweden}
\altaffiltext{3}{Laboratoire d'Astrophysique de Marseille, UMR 6610,
  CNRS-Universit\'e de Provence, 38 rue Fr\'ed\'eric Joliot-Curie, 13388
  Marseille Cedex 13, France}
\altaffiltext{4}{Institute of Theoretical Astrophysics, University of Oslo, 
  P.O. Box 1029, Blindern, N-0315 Oslo, Norway}
\altaffiltext{5}{Centre of Mathematics for Applications, University of Oslo,
  P.O. Box 1029, Blindern, N-0315 Oslo, Norway}
\altaffiltext{6}{Space Telescope Science Institute, 3700 San Martin Drive,
  Baltimore, MD 21218, USA} 
\altaffiltext{7}{Dark Cosmology Centre, Niels Bohr Institute, University
  of Copenhagen, Juliane Maries Vej 30, DK-2100 Copenhagen, Denmark}
\altaffiltext{8}{Department of Physics and Astronomy, University of
  Victoria, PO Box 3055 STN CSC, Victoria BC V8T 1M8, Canada}
\altaffiltext{9}{Australian Astronomical Observatory, Epping, NSW 1710,
  Australia}
\altaffiltext{10}{Physikalisches Institut Universit\"at Bonn, Nussallee 12, 
  53115 Bonn, Germany}
\altaffiltext{11}{CRAL, Observatoire de Lyon, Universit\'e Lyon 1, 9 Avenue
  Ch. Andr\'e, 69561 Saint Genis Laval Cedex, France}
\altaffiltext{12}{Department of Astronomy, Stockholm University,
  Albaova University Center, SE~106-92 Stockholm, Sweden}
\altaffiltext{13}{CENTRA-Centro Multidisciplinar de Astrof\'{\i}sica,
  IST, Avenida Rovisco Pais, 1049-001 Lisboa, Portugal}
\email{rahman@fysik.su.se}



\begin{abstract}
  \noindent Our ability to study the most remote supernova explosions, crucial
  for the understanding of the evolution of the high-redshift universe
  and its expansion rate, is limited by the light collection
  capabilities of telescopes. However, nature offers unique
  opportunities to look beyond the range within reach of our unaided
  instruments thanks to the light-focusing power of massive galaxy
  clusters.  Here we report on the discovery of one of the most
  distant supernovae ever found, at redshift, $z=1.703$.  Due to a
  lensing magnification factor of $4.3\pm0.3$, we are able to measure
  a lightcurve of the supernova, as well as spectroscopic features of
  the host galaxy with a precision comparable to what will otherwise
  only be possible with future generation telescopes.
\end{abstract}





\section{Introduction}
Supernovae (\sne), exploding stars at the end of their life cycles,
have several astrophysical and cosmological
applications. Core-collapse \sne trace the star formation history
\citep{2004ApJ...613..189D,2009A&A...499..653B} while the standard
candle property of Type Ia \sne (\sneia) can be used for probing the
expansion history of the universe \citep[see, e.g.,][and references
therein]{2010ApJ...716..712A,2011ApJ...737..102S}. In particular, it
is desirable to study \sne in the distant universe. Our ability to do
this is currently limited by the light collecting power of existing
telescopes.



Massive galaxy clusters, $M\gsim10^{14}\,M_{\odot}$, act as powerful
gravitational telescopes, providing the capability to push
observations to higher redshifts \citep{2004ApJ...607..697K}.
For background limited observations, the magnification, $\mu$,
corresponds to a gain factor in exposure length (or mirror area) of
$\mu^2$, which is of particular importance for increasing the depth in
the near-IR. For example, in the {\it J}-band \citep{esoskynir} the
atmosphere is $\sim3$~mag brighter than in {\it I}
\citep{esoskyoptical}, and yet another $\sim2.5$~mag brighter in the
{\it Ks}-band.



The feasibility of detecting high-$z$ \sne along the line of
sight of massive clusters \citep{2003A&A...405..859G} was first
explored by our team using the Infrared Spectrometer And Array Camera
(ISAAC) at European Southern Observatory's (ESO) Very Large Telescope
(VLT), although it has also been discussed in previous work
\citep{1988ApJ...335L...9K,1998MNRAS.296..763K,2000MNRAS.319..549S,2002MNRAS.332...37G}.
Traditional SN searches are done at optical wavelengths where \sne
typically emit most of their light. However, at high redshift the
optical region is shifted to the near-IR which is why we chose to
carry out our survey at these wavelengths. In
\citet{2009A&A...507...61S,2009A&A...507...71G} we reported the
discovery of a highly magnified \sn, yet severely dimmed by
dust, at $z\sim0.6$ behind one of the best studied galaxy clusters,
A1689. In this work we report on the discovery of one of the most
distant \sne ever found, at redshift, $z=1.703$, from our
$\sim10$\,hr VLT survey of the same cluster.

\section{Observations}
We have used the High Acuity Wide field {\it K}-band Imager
\citep[HAWK-I;][]{2004SPIE.5492.1763P,2006SPIE.6269E..29C,2008A&A...491..941K,2011Msngr.144....9S}
camera on the VLT
to monitor galaxies behind the galaxy cluster A1689
(R.A.=13:11:30, decl.=$-$01:20:28 J2000 at redshift, $z=0.187$) from
2008 December to 2009 July, with observations separated by
approximately one month. Each visit consisted of $\sim2$~hr long
integrations in the $J$-band ($\sim1170$--$1340\,$nm), with seeing
conditions of $0\farcs60$--$0\farcs90$~FWHM. At the same time, we
carried out a similar programme using The Andalucia Faint Object
Spectrograph and Camera (ALFOSC) at the Nordic Optical Telescope (NOT)
in the {\it i}-band ($\sim690$--$850\,$nm). Images from different
epochs were aligned, seeing matched, and subtracted in order to find
transient objects using the method described in
\citet{2008A&A...486..375A}. The most distant transient found in the
survey was first detected on UT\,2009~June~5 in the optical NOT {\it
  i}-band data and was confirmed with a detection in the HAWK-I {\it
  J}-band data two days later.  Follow-up photometry was obtained for
the three following months until the target disappeared behind the
Sun.  All visits are listed in Table~\ref{tb:observations}.




\section{The transient at $z=1.703$}
\begin{figure}[tbp]
  \centering
  \includegraphics[width=0.7\textwidth]{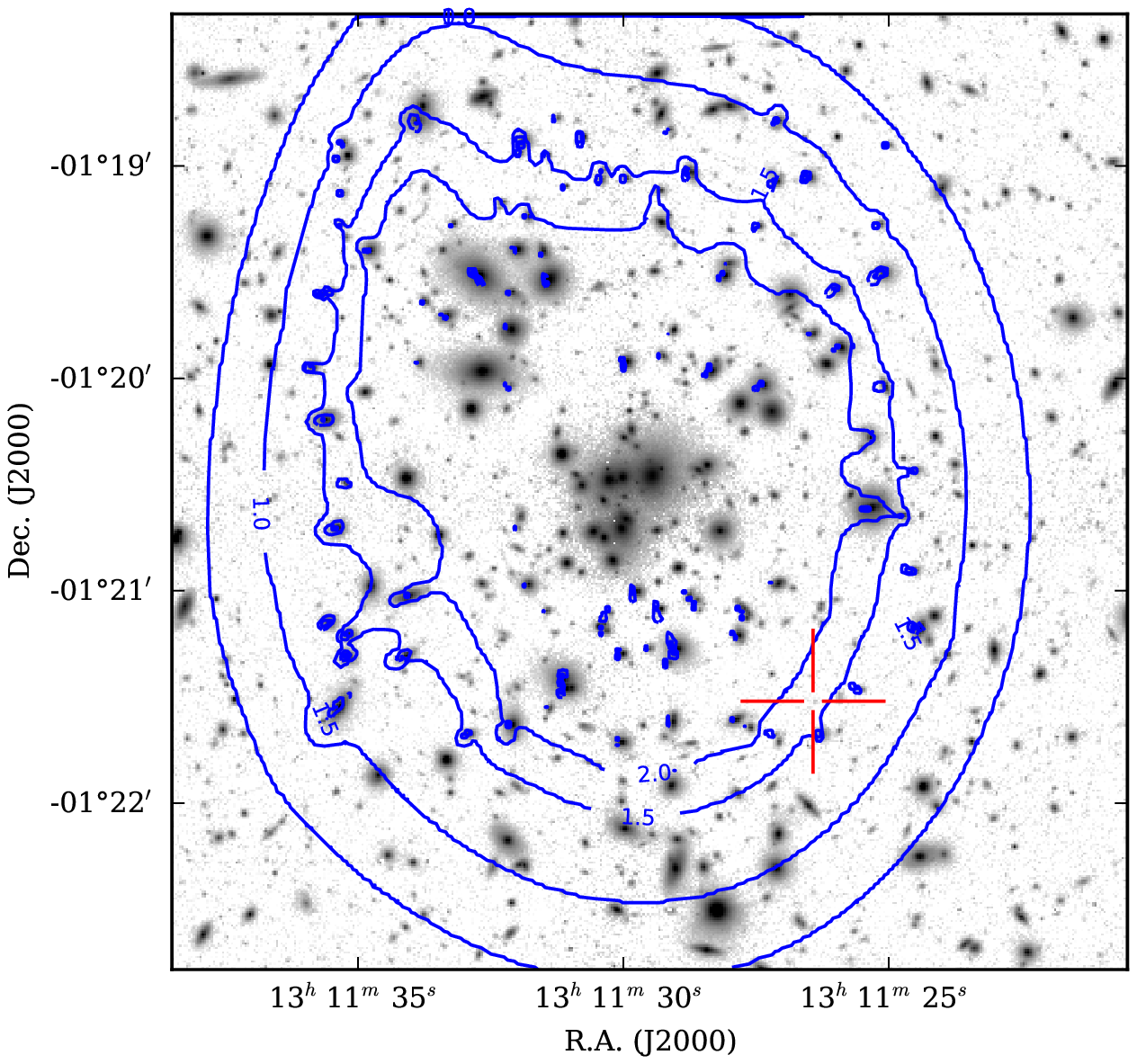}
  \caption{A1689, R.A.=13:11:30, decl.=$-$01:20:28 (J2000).  A
    cropped $4.'5$ field of the stacked {\em J}-band HAWK-I image from
    the survey centered on the cluster. The supernova position,
    R.A.=13:11:26.426 and decl.=$-$01:21:31.21 (J2000), is indicated by the
    red cross hair, while blue solid lines are the iso-magnification
    contours in units of magnitudes, $\Delta m_\mu=2.5\log_{10}\mu$,
    for a source at $z=1.7$ derived from the strong lensing
    analysis. For clarity, the contours have been omitted from the
    cluster center.
    \label{fig:pos}}
\end{figure}

\begin{figure}[p]
  \centering
  \includegraphics[width=0.64\textwidth]{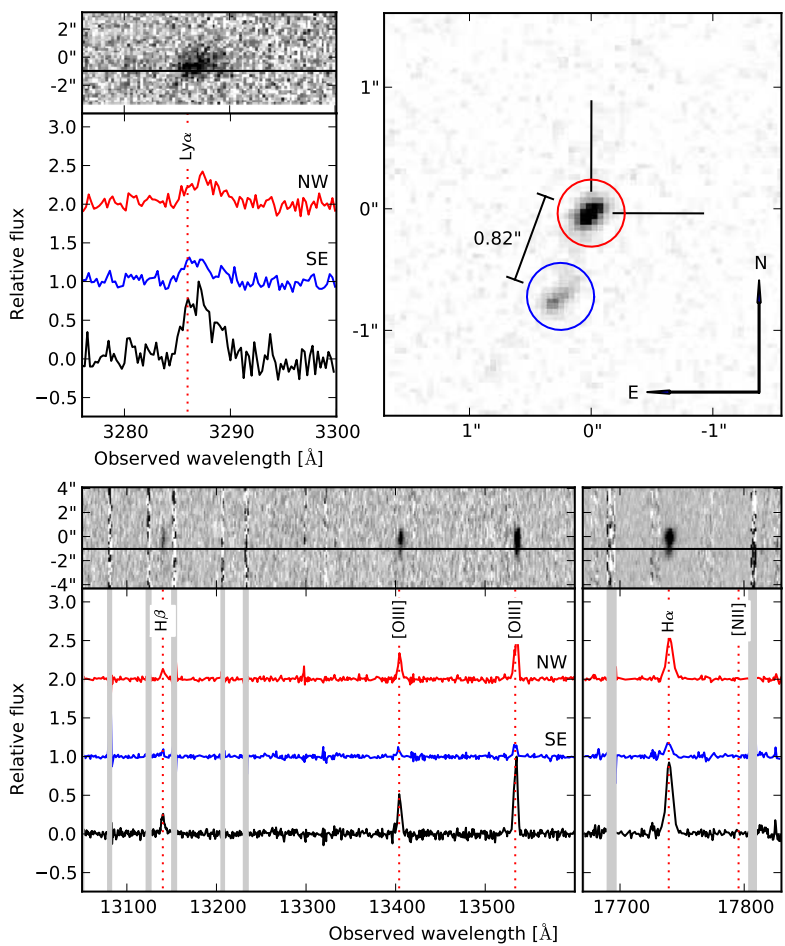}
  \caption{Spectrum of the supernova host galaxy. 
    Top right: the host galaxy from ACS archive images in the
    F775W band. The supernova position has been marked by the black
    cross hair, and the two components of the host are marked by
    circles. At the given redshift, $0\farcs82$ corresponds to $7\,$kpc.
    Top left and bottom: host spectra from VLT X-SHOOTER 
    showing the identified emission lines.  Atmospheric lines are marked by
    the shaded regions. No continuum was detected.  Each panel shows
    the two-dimensional spectrum where the spatial scale is expressed 
    relative to the northern host component and the distance between the 
    two has been marked.  The top and middle one-dimensional spectra show 
    the northern and southern components respectively, while the bottom
    is the sum of the two.
    \label{fig:hostspec}}
\end{figure}

\newcommand{\lya}{\mathrm{Ly}\alpha\,\lambda1216}
\newcommand{\oii}{[\mathrm{O\textsc{ii}}]\,\lambda\lambda3726,3729}
\newcommand{\oiii}{[\mathrm{O\textsc{iii}}]\,\lambda\lambda4959,5007}
\newcommand{\ha}{\mathrm{H}\alpha\,\lambda6563}
\newcommand{\hb}{\mathrm{H}\beta\,\lambda4861}
\newcommand{\hg}{\mathrm{H}\gamma\,\lambda4340}
\newcommand{\nii}{[\mathrm{N\textsc{ii}}]\,\lambda6583}
\newcommand{\oiiihb}{[\mathrm{O\textsc{iii}}]\,\lambda5007/\mathrm{H}\beta}
\newcommand{\niiha}{[\mathrm{N\textsc{ii}}]\,\lambda6584/\mathrm{H}\alpha}
\newcommand{\bpty}{\log_{10}\left(\oiiihb\right)}
\newcommand{\bptx}{\log_{10}\left(\niiha\right)}
Figure~\ref{fig:pos} shows the position of the transient on the sky,
while a zoomed view of the host galaxy can be seen in the top-right
panel of Figure~\ref{fig:hostspec}.  The transient was detected
$60\arcsec$\,W and $50\arcsec$\,S of the cluster center, located
$0\farcs04\pm0\farcs03$ from the core of the northwestern of the
two knots that appear to be part of the same system with a separation
of $\sim7\,$kpc. The host galaxy redshift was initially estimated to
$z=1.65\pm0.10$ from multi-band photometry. One major advantage of
searching for \sne in lensed galaxies is that it also allows for
accurately studying the host environment. A 1~hr VLT XSHOOTER
\citep{2006SPIE.6269E..98D} spectrum (\citeauthor[][{\it in
  preparation}]{watson:2012}) of the host galaxy, shown in
Figure~\ref{fig:hostspec}, was obtained on 2010~April~23. The redshift
could accurately be determined to $z=1.703$ from the identified
$\lya$, $\hb$, $\oiii$ and $\ha$ lines. We also detected a weak $\hg$
line, but no $\oii$, nor $\nii$. Using the method from
\citet{2004MNRAS.348L..59P}, we derived an upper limit on the
metallicity of $12 + \log(\mathrm{O/H})\lesssim8.1$. Further, the line
ratios $\oiiihb$ and $\niiha$ can be used to separate star-forming
galaxies from those hosting active galactic nuclei
\citep[AGNs;][]{1981PASP...93....5B,2003MNRAS.346.1055K}, e.g.,  no
narrow-emission line AGNs have been observed with $\bptx<-0.7$
\citep{2006MNRAS.372..961K}.  Due to the absence of $\nii$ in our
spectra, we can put an upper limit on the latter ratio of
$\bptx\lesssim-1.3$.

\begin{figure}[tbp]
  \centering
  \includegraphics[width=\textwidth]{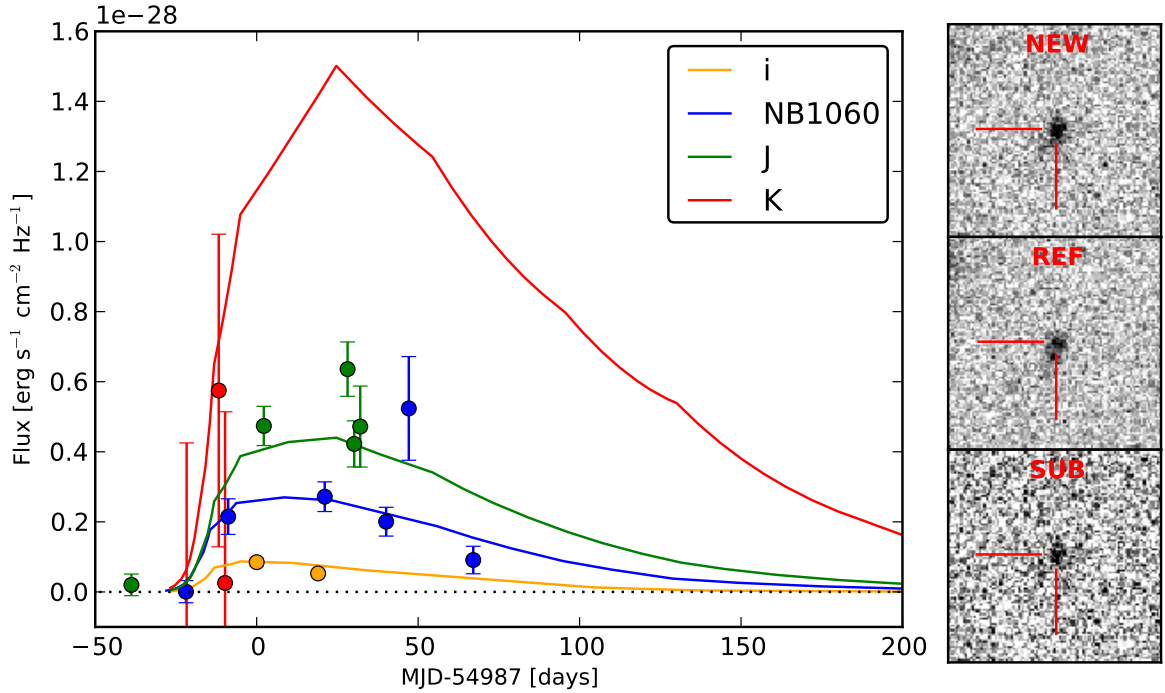}
  \caption{Supernova light curve.
    Light curve of the supernova at $z=1.703$. The solid lines show the
    best-fitted template, a Type~IIn SN, from Peter Nugent based on
    SN~1999el \citep{2002ApJ...573..144D}.  The time of maximum and 
    the amplitude of the template have been fitted. The right panel
    shows the subtraction (SUB) of the $J$-band HAWK-I images from 
    2009 June to July (NEW), and 2009~January (REF).
    \label{fig:lc}}
\end{figure}

While the transient was active, the galaxy cluster A1689 was
observed in {\it Ks} ($\sim1960$--$2400\,$nm) and the narrow-band
filter {\it NB1060} ($\sim1055$--$1070\,$nm) with HAWK-I in ESO
programme 181.A-0485. Combining the observations from all three HAWK-I
filters and the NOT observations allowed us to build the four-band
light curve shown in Figure~\ref{fig:lc}. The transient is most likely
a supernova based on the fact that (1)~no transient activity has been
observed, neither in archival data, nor in the continuous optical
follow-up $1.5\,$years after the discovery, (2)~the observed
light curve, and the absolute magnitude are consistent with the
expectations for an \sn at the redshift of the host galaxy, and
(3)~the host galaxy spectrum is inconsistent with an AGN as
described above.

The significant detection in the rest-frame ultraviolet (UV), observed
$i$~band, excludes a thermonuclear supernova (SN~Ia), and it can be
concluded that the observed transient is most likely a core-collapse
\sn. It should be pointed out that the rest-frame optical lightcurve
(the HAWK-I bands) is also consistent with an SN~Ia, emphasizing the
importance of follow-up over a broad wavelength range when no
spectroscopic confirmation is available.

There are only very few \sne that have been observed in
rest-frame ultraviolet, in particular with coverage in the full
wavelength range that was observed here. In order to constrain the
nature of the \sn further, we carried out template fitting for
different \sn types. The template that matches our data best is
shown in Figure~\ref{fig:lc}. This corresponds to Peter
Nugent's\footnote{\texttt{http://supernova.lbl.gov/nugent/$\sim$nugent\_templates.html}}
Type~IIn~supernova (SN~IIn) based on SN~1999el
\citep{2002ApJ...573..144D}, where the fitted parameters are the time
of maximum light together with the flux normalization. The color of
the SN is consistent with the used template. It can be debated whether
it is appropriate to fit the narrowband {\em NB1060} data on equal
basis with the other bands, since diversity between individual \sne is
certainly expected in narrow wavelength regions. However, refitting
the lightcurve omitting this filter has only a minor impact on the
fitted parameters and does not change the SN typing.
We have no significant detection of spectroscopic features that can
only be associated with the transient.  Note, however, that the
X-SHOOTER spectrum was obtained $\sim120$~days past maximum light in
the rest frame.

The mass distribution model \citep{2007ApJ...668..643L} for A1689 is
well determined thanks to the large number of constraints based on
deep archival multi-band Hubble Space Telescope ({\it HST})
observations. It is based on weak and strong lensing analysis of
background galaxies and uses 34 multiply imaged systems, of which 24
have spectroscopic redshifts. Using the model, the cluster
magnification at the position of the transient for redshift $z=1.703$
is estimated to be $\Delta m_\mu=1.58\pm0.07$~mag
\citep[consistent with the updated model used
in][]{2011arXiv1109.6351R} where $\Delta m_\mu=2.5\log_{10}\mu$. Given
the lensing magnification, and assuming standard cosmological
parameters ($h,\Omega_M.\Omega_\Lambda$)=($0.74,0.27,0.73$), the
absolute $V$-band magnitude is %
$M_V=-19.56\pm0.06\mathrm{(phot)}\pm0.07\mathrm{(lens)}$. This places
the magnitude of the \sn roughly one magnitude brighter than the
mean absolute magnitude of SNe~IIn in the local universe
\citep{2002AJ....123..745R,2010arXiv1010.2689K}. Note, however, that
several SNe~II much brighter than this have been found in the past
\citep{2008ApJ...686..467S}.

\section{Discussion}
%
%
We have demonstrated a relatively inexpensive technique for opening up
a high-redshift window for finding \sne, that allows for
photometric and, for high magnifications, spectroscopic follow-up, in
the \emph{rest-frame optical} of both \sne and their hosts.  In
our $\sim10$\,h lensing aided survey we found one \sn at
$z=1.703$ which is consistent with the expected number of 0.3~\sneia
and 0.8~core collapse at $z\geq1.5$.  A full rate analysis including
discoveries at lower redshifts is the topic for an upcoming paper.

A handful of \sne from other surveys have been discovered at
redshifts comparable to what is reported here, but they have all been
found from much more extensive programs.  For example, SN\,1997ff at
$z\sim1.7\pm0.1$ \citep{2001ApJ...560...49R} was found by repeated
{\it HST} observations of the Hubble Deep Field-North,
while \cite{2009Natur.460..237C} reported the discovery of three
$z\sim2$ Type IIn \sne in seasonal-stacked images from the Supernova
Legacy Survey.  Further, repeated optical imaging of the Subaru Deep
Field from 2002 to 2008 has resulted in 10 single-epoch discovered
\sne in the range $1.5<z<2.0$
\citep{2007MNRAS.382.1169P,2011arXiv1102.0005G}.

\begin{figure}[tbp]
  \centering
  \includegraphics[width=\textwidth]{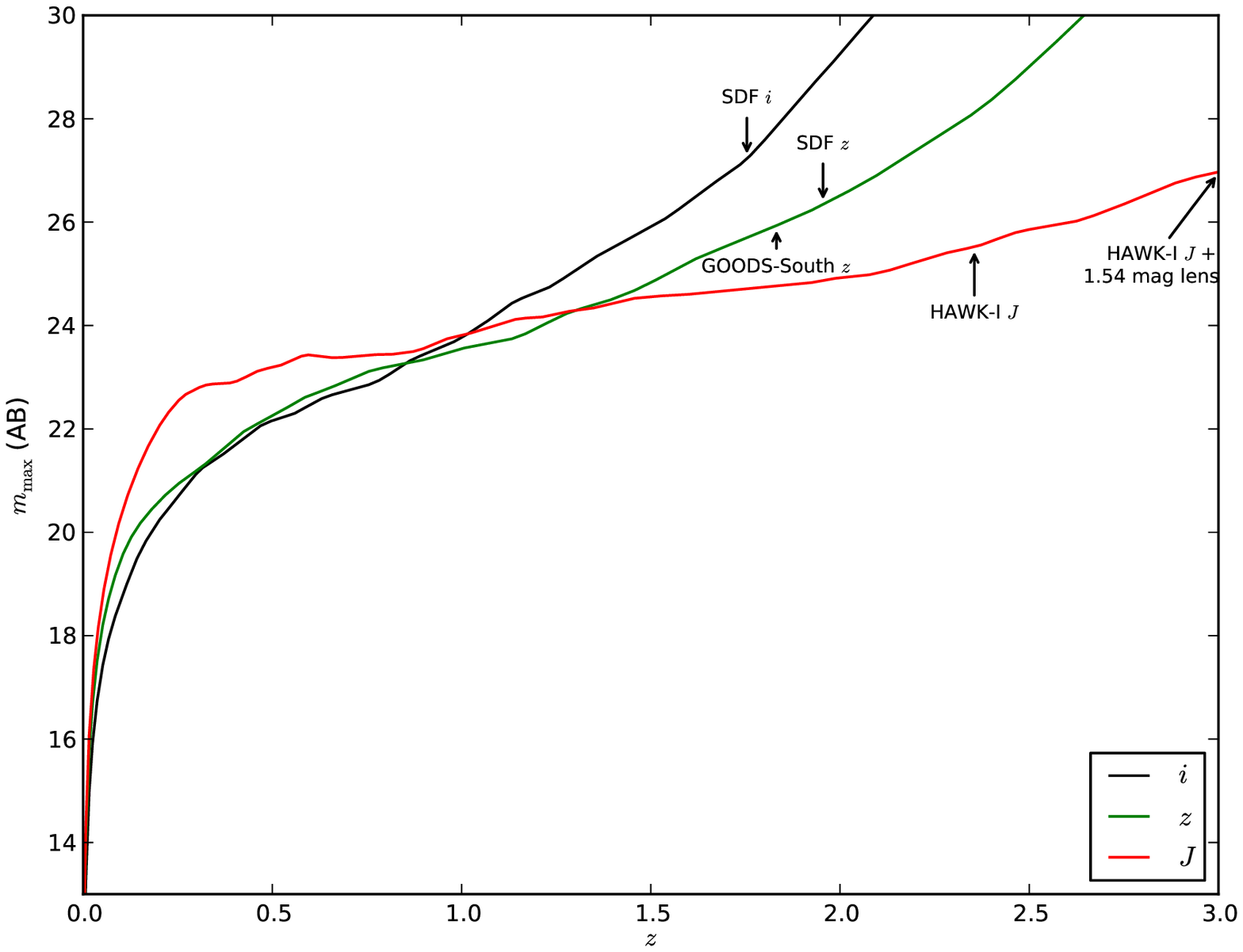}

  %
  %
  \caption{%
    \snia peak magnitude in different passbands for increasing
    redshifts
    assuming an absolute magnitude of $M_B=-19.2$ for
    $H_0=74\,$km\,s$^{-1}$ \citet{2011ApJ...730..119R}. The arrows
    indicate the redshifts that were reached for a $5\sigma$ detection
    for the SDF survey as described in \citet{2011arXiv1102.0005G},
    the {\it HST} GOODS-South survey \citep{2010ApJ...724..425D}, and our
    own survey with HAWK-I.  For the latter we also show the limit
    that can be reached assuming the same lensing magnification as for
    the \sn discussed in this Letter. \label{fig:surveydepth}}
\end{figure}

Although wide-field optical surveys may be useful for finding high-$z$
\emph{UV-bright} \sne they are less efficient for finding, e.g., \sneia
with their emission peaking in the rest-frame {\it B} band. Carrying
out observations in the near-IR is essential for finding these.  To
illustrate this, the expected magnitude of \sneia for different pass
bands as a function of redshift is shown in
Figure~\ref{fig:surveydepth} together with the magnitude limits for
the deepest \sn surveys to date.
From this it can be concluded that it is almost impossible to find
\sneia at $z\gsim2$ in an optical survey.  Further, even if such
objects are found, using them for probing cosmology requires at the
very minimum photometric follow-up in one additional filter, and
preferably spectroscopic confirmation, which is very difficult with
existing ground-based telescopes in an unlensed scenario.


Exploiting this technique for several comparable clusters and over an
extended period of time would have important implications for the
study of high-redshift star formation, which is closely linked to the
rate of core collapse \sne, the progenitor scenario of SNe~Ia, and of
course cosmology with SN~Ia, most specifically the nature of dark
energy. Although these are key goals for future instruments such as
the {\it James Webb Space Telescope}, the successor of the {\it HST},
or ground based extremely large telescopes ($>25$\,m) we can start
addressing them already today thanks to Nature's own gravitational
telescopes. Long-term monitoring of massive clusters will also yield
strongly lensed, multiply imaged \sn images, expected at a rate
of about once every five years per cluster, for which a time delay
will provide independent information about the cosmological distance
scale to a precision of a few percent.



\acknowledgments The authors thank Claes Fransson for many useful
discussions. The work is based, in part, on observations obtained at
the \facility{ESO Paranal Observatory} (ESO programmes
082.A-0431;0.83.A-0398, PI:~A.~Goobar, 085.A-0909, PI:~D.~Watson
181.A-0485; PI: J.~G.~Cuby).  It is also based, in part, on
observations obtained at the \facility{Nordic Optical Telescope} (NOT
programme P39-011, PI:~A.~Goobar) with ALFOSC, which is provided by
the Instituto de Astrofisica de Andalucia (IAA) under a joint
agreement with the University of Copenhagen and NOTSA. The Dark
Cosmology Centre is supported by the Danish National Research
Foundation.

\clearpage




\clearpage





\clearpage

\begin{table}[tbp]
  \caption{VLT and NOT observations of A1689 during the 2009 survey. 
    The late NB1060 observations taken during consecutive nights were 
    stacked. The AB magnitudes of all SN points with $>2\sigma$ detection
    are listed in the last column.
    \label{tb:observations}}
  \vspace{2em}
  \centering
  \begin{tabular}{l c l c r r}
    \hline
    Instrument &    Filter    & Date (UT)  & MJD     & 
    \multicolumn{1}{c}{Exp. time [s] }
    & \multicolumn{1}{c}{SN mag (AB)}\\
    \hline
%
%
    HAWK-I & {\em J}      & 2008 Dec 30 & 54831.3 &  3420 & \\
    HAWK-I & {\em J}      & 2008 Dec 31 & 54832.4 &  2400 & \\
    HAWK-I & {\em J}      & 2009 Jan 2 & 54834.4 &  2400 & \\
    HAWK-I & {\em J}      & 2009 Jan 30 & 54862.3 &  7200 & \\
    HAWK-I & {\em J}      & 2009 Mar 1 & 54892.3 &  4800 & \\
    HAWK-I & {\em J}      & 2009 Mar 23 & 54914.4 &  4800 & \\
    HAWK-I & {\em J}      & 2009 Mar 25 & 54916.2 &  4800 & \\
    HAWK-I & {\em Ks}     & 2009 Mar 27 & 54918.3 &  2800 & \\
    HAWK-I & {\em NB1060} & 2009 Mar 28 & 54919.2 &  3300 & \\
    HAWK-I & {\em NB1060} & 2009 Apr 1 & 54923.3 &  6600 & \\
    HAWK-I & {\em NB1060} & 2009 Apr 5 & 54927.2 &   600 & \\
    HAWK-I & {\em Ks}     & 2009 Apr 5 & 54927.3 &  2970 & \\
    ALFOSC & {\em i}      & 2009 Apr 21 & 54943.0 &   900 & \\
    HAWK-I & {\em Ks}     & 2009 Apr 26 & 54948.3 &  2970 & \\
    HAWK-I & {\em NB1060} & 2009 Apr 27 & 54949.0 &  3300 & \\
    HAWK-I & {\em J}      & 2009 Apr 27 & 54949.2 &  7200 & \\
    HAWK-I & {\em NB1060} & 2009 May 14 & 54966.1 & 16500 & \\
    HAWK-I & {\em Ks}     & 2009 May 14 & 54966.2 &  2970 & \\
    HAWK-I & {\em NB1060} & 2009 May 17 & 54969.0 &  6600 & \\
    HAWK-I & {\em Ks}     & 2009 May 24 & 54976.2 &  2970 & \\
    HAWK-I & {\em Ks}     & 2009 May 26 & 54978.1 &  2970 & \\
    HAWK-I & {\em NB1060} & 2009 May 27 & 54979.1 &  3600 & 25.36\,(0.27)\\
    ALFOSC & {\em i}      & 2009 Jun  5 & 54987.9 &   900 & 24.07\,(0.15)\\
    HAWK-I & {\em J}      & 2009 Jun 7 & 54990.1 &  2400 & 26.06\,(0.13)\\
    ALFOSC & {\em i}      & 2009 Jun 24 & 55006.9 &   900 & 24.60\,(0.18)\\
    HAWK-I & {\em NB1060} & 2009 Jun 26--27 & 55009.0 &  9900 &
    25.11\,(0.17)\\
    HAWK-I & {\em J}      & 2009 Jul 3 & 55016.1 &  2400 & 25.74\,(0.13)\\
    HAWK-I & {\em J}      & 2009 Jul 5 & 55018.1 &  2400 & 26.19\,(0.17)\\
    HAWK-I & {\em J}      & 2009 Jul 7 & 55020.0 &  2400 & 26.07\,(0.27)\\
    HAWK-I & {\em NB1060} & 2009 Jul 15 & 55028.0 &  7200 & 25.44\,(0.22)\\
    HAWK-I & {\em NB1060} & 2009 Jul 22 & 55035.0 &  3300 & 24.39\,(0.31)\\ 
    HAWK-I & {\em NB1060} & 2009 Aug 11--13 & 55055.0 &  10200 &
    26.29\,(0.47)\\
    \hline
  \end{tabular}
\end{table}

\end{document}